\documentclass{elsart}

\usepackage{lineno,hyperref}
\modulolinenumbers[5]

\usepackage{graphics}
\usepackage{epsfig}

\usepackage{amssymb}

\usepackage{soul}
\usepackage{xcolor}
\begin{document}
\title{Study of medium modified jet shape observables in Pb-Pb collisions at $\sqrt{s_{NN}}$~=~2.76 TeV using EPOS and JEWEL event generators}
\begin{frontmatter}
\author{Sumit Kumar Saha$^1$}, 
\author{Debojit Sarkar $^{2,3,4}$ \corauthref{cor1}}, 
\corauth[cor1]{debojit03564@gmail.com}
%\ead{debojit03564@gmail.com}
\author{Subhasis Chattopadhyay$^1$ },
\author{Ashik Ikbal Sheikh$^1$}
\author {and}
\author{Sidharth Kumar Prasad$^3$}
\address{$^1$ Variable Energy Cyclotron Centre, HBNI, 1/AF-Bidhannagar, Kolkata-700064, India}
\address{$^2$ Wayne State University, 666 W. Hancock, Detroit, MI 48201, USA}
\address{$^3$ CAPSS, Bose Institute, Block EN, Sector 5, Kolkata 700091, India.}
\address{$^4$ {Laboratori Nazionali di Frascati, INFN, Frascati, Italy}}

\begin{abstract}
The jet-medium interaction in  high energy heavy ion collisions is an important phenomena to characterize the hot and dense medium produced in such collisions. The study of medium-induced modifications to the substructure of inclusive charged jets indicates a redistribution of energy inside the jet cone and provides insight into the energy loss mechanisms of jets in the medium. We investigate the in-medium modification to two jet shape observables i.e., the differential jet shape ($\rho$(r)) and the angularity (g)  in the most central $Pb-Pb$ collisions at $\sqrt{s_{NN}} ~=~ 2.76 $ TeV using two commonly used event generators i.e., JEWEL (recoil OFF) and EPOS-3 in the jet-p$_T$ range of 20-40 GeV/c.  JEWEL with recoil OFF has been used primarily as a reference system as that has been found to explain the global jet observables satisfactorily but lacks in jet-shape variables at higher jet-radii. EPOS-3 that explains the bulk properties in such collisions quite well takes into account a hydrodynamically evolving bulk matter, jets and hard-soft interactions. A comparison between the results from these models shows that while JEWEL (recoil OFF) does not explain the distribution of lost energy at higher radii with respect to the jet-axis, EPOS-3 explains the effect quite well. However, in EPOS-3, the partonic energy loss mechanism and secondary hard-soft interactions during hadronization and hadronic cascade phase are different from the conventional jet energy loss models. The current study can, therefore, provide important new insights on mechanisms regarding the modeling of the medium and hard-soft interactions in heavy ion collisions. 
\end{abstract}
\end{frontmatter}

% main text
\maketitle

\section{Introduction}

In high energy heavy ion collisions at the RHIC and the LHC energies, as per available data, a medium with partonic degrees of freedom is formed ~\cite{sergei_flow}. The transition from a confined hadronic phase to a deconfined partonic phase has been concluded to be a cross-over \cite{Susskind:1979up}. A wide range of observables measured in Au-Au/Pb-Pb collisions and their comparison with the reference systems like pp collisions shed light on various properties of the medium. One of the early observables in this field of study that probed the gluon-density of the medium had been the fragments of highly energetic  partons in terms of high p$_T$ leading particles  and additionally reconstruction of full jets at the RHIC and LHC energies \cite{jet_review, br, ph, phn, str, al}. Initially, global jet-observables in the form of suppression of both the leading particles and jets, commonly known as jet-quenching have been used to probe the gluon-density of the medium. \cite{Bjorken, jet_quench_1, jet_quench_3}.

Now, energy loss of partons by radiation or collision is expected to modify the fragmentation function of the incoming partons. It is expected that during the process of energy loss and hadronization, the internal structure of jet also undergoes modification. Measurements of observables like transverse spread of energy and momentum of the jet fragments in central heavy ion collisions and it's comparison to pp collisions lead to a conclusion that the core of the jet gets harder and the periphery gets extended to a larger radii with softer fragments due to jet-medium interaction \cite{ATLAS:2012ina}. Observables like energy asymmetry (A$_j$) and variation of particle density inside a jet cone with radial distance have been used to understand the energy distribution within jets after quenching \cite{cms1,cms2}.

As per theoretical descriptions of jet quenching \cite{Bjorken, Aamodt:2010jd, Chatrchyan:2011sx, Betz:2012hv, rhic, lhc}, high energetic partons suffer energy loss due to interactions with thermal partons in the medium and these scattered medium partons can have an effect on the final jet-shape parameters~\cite{jet_shape_1, jet_shape_2, jet_shape_3}. Studies are ongoing using various models for estimating the effect of jet-quenching on jet-shape observables. These models mostly describe the global jet observables like R$_{AA}$ quite well \cite{dijet_asym_1, dijet_asym_2, Krofcheck:2013dua}. One model that has been extensively used in such studies is JEWEL \cite{jewel} that describes the global jet properties quite satisfactorily at  the LHC energy. The model as discussed in the next section in detail, does not simulate the heavy ion collisions as a whole. Rather it uses perturbative QCD to describe the interaction of the hard scattered partons with an ensemble of partons whose phase space distribution and flavor composition is provided by simple Bjorken model \cite{JEWEL_main}. In JEWEL, interaction of the shower partons with the medium partons  can be treated in two modes i.e., recoil OFF and recoil ON. In the recoil OFF version, the recoiled medium partons do not take part in further processes towards hadronization. The recoil-ON version, on the other hand propagates the effect of those recoiled partons to the final observables. Analysis of results from the recoil-ON version however faces the challenges of background subtraction which is necessary for comparison with the experimental results \cite{dijet_asym_2}. Even though extensive efforts have been made to develop background subtraction methods, they have their own limitations. The recoil-OFF version can however be compared with experimental results directly.
%Analyzing the jet-shape parameters therefore will help to understand the underlying interaction of jets with the medium.
Experimentally, for very high p$_T$ jets($>$ 100 GeV/c), the background reduces drastically mainly due to higher kinematic cuts applied on the fragments for jet reconstruction \cite{Connors:2017ptx}. But for jets with relatively lower p$_T$, background effect is more prominent along with other softer contributions from the medium. This issue is similar for both data and event-generators simulating jet-medium interactions. At a jet p$_T$ range of say 20-40 GeV/c, the absence of full event simulation in JEWEL coupled to the uncertainties in various background subtraction methods together make the predictions of jet-shape observables with recoil-ON version even more complicated. In this study, we have taken JEWEL with recoil-OFF and studied the jet p$_T$ range of 20-40 GeV/c with R~=~0.2 and 0.3 as a reference.

In addition to the results from JEWEL taken as reference, we have studied the same set of jet-observables using another model named EPOS-3  \cite{EPOS_main, Porteboeuf:2010um} that explains the bulk  observables in high energy heavy ion collisions quite satisfactorily \cite{epos_model_descrip, lambda_suppression_ALICE, epos_PbPb_lambdakshort_enhance, EPOS_ridge_pPb}. However the model has not been tested well for hard probes. EPOS-3, a full event generator and described in the next section in detail considers the collision zone consisting of two regions called core and corona. The core undergoes 3+1 D event by event hydrodynamic evolution and explains observables like flow, particle production at low p$_T$ quite well. The corona region, on the other hand consists of jets and implements the high p$_T$ phenomena. The simplistic implementation of the partonic energy loss in EPOS-3 is different from the conventional jet-energy loss models \cite{EPOS_main}. In addition, a modified hadronization procedure through recombination of the corona string segments produced within the fluid freeze-out hypersurface with the core string segments has been found to be essential in explaining the $R_{AA}$ upto p$_{T}$ $\approx$ 20 GeV/c  in heavy ion collisions. In EPOS-3, the probability of formation of jet-hadrons inside the fluid freezeout surface is considerably high upto p$_{T}$ $\approx$ 20 GeV/c  and these jet-hadrons also have a large probability of re-scattering with the soft hadrons from freeze-out \cite{EPOS_main}. These secondary hard-soft interactions can modify the distribution of jet particles inside the jet cone. Hence, the comparison of the jet shapes obtained using EPOS-3 and JEWEL will shed some light on the different physics processes which can potentially explain the jet-shape broadening in heavy ion collisions.

In this work, we reconstruct jets from two models with two R values of 0.2 and 0.3 with a jet p$_T$ range of 20-40 GeV/c. This low-intermediate jet p$_T$ range, where medium induced modifications to the jets are stronger, consists of jet constituents having p$_{T}$ $<$ 20 GeV/c and suitable for investigating the effect of secondary hard-soft interactions in EPOS-3 as described above. We have studied a set of jet-shape observables as described in section-III. Higher R values access higher transverse region and thereby explores the lost jet energy distribution in greater detail.

Main motivation of this work can be listed as (a) sensitivity of JEWEL on jet-shape at lower p$_T$ region without considering the recoiled partons. The possible difference in the pattern compared to the experimental data may help to better understand the effect of the recoiled medium partons on the jet-shape observables. (b) Study the jet-shape observables in the same p$_T$ range using EPOS-3 that includes a simplistic partonic energy loss mechanism and secondary hard-soft interactions during hadronization and the hadronic cascade phase.

This article is organised as follows, in the next section we provide further details on two models i.e. JEWEL and EPOS-3 in the context of jet quenching, in section-III, we discuss the observables and the analysis method adopted here. Section IV presents the results and the discussions.

\section{Event generators: JEWEL and EPOS}

EPOS-3  is a parton based model with flux tube initial conditions  \cite{Werner:2013tya, Werner:2010ss, Werner:2013ipa, Drescher:2000ha}. Initially, the partons undergo multiple scatterings and the final state partonic system consists of mainly longitudinal color flux tubes (strings) carrying transverse momentum of the hard scattered partons in the transverse direction  \cite{epos_model_descrip, epos_core_corona_sep}. Depending on the partonic energy loss scheme as described in \cite{EPOS_main} and local string density, these strings will eventually form both the core (bulk) and the corona (jet). The low momentum strings in the high density area undergo hydrodynamic evolution and form the bulk. Whereas, the highly energetic strings in the low density area form the corona (jet) following Schwinger mechanism. In EPOS, the partonic energy loss scheme doesn't involve the interaction of jets with hydrodynamic fields and it depends on the initial geometric size of the fireball and density of the string segments along with some other parameters. The secondary hard-soft interactions take place during the hadronization  and the hadronic cascade phase \cite{EPOS_main}. For example, the intermediate $p_{T}$ corona string segments have significant probability of forming inside the fluid freeze-out surface and these segments may pick up partons from the thermal matter rather than creating them via usual Schwinger mechanism \cite{EPOS_main, EPOS_ridge_pPb}. In addition, these jet hadrons will also suffer hadronic interactions with the soft hadrons from the freeze out and may further get re-distributed within the jet cone. This approach has been found to be essential in describing several experimental features such as the nuclear modification factor ($R_{AA}$) \cite{EPOS_main, lambda_suppression_ALICE}, baryon to meson enhancement at intermediate $p_{T}$  \cite{epos_PbPb_lambdakshort_enhance}, elliptic flow at higher $p_{T}$ \cite{EPOS_main} etc in heavy ion collisions.  In this work, the jet-shape will reflect the effect of the hard-soft interactions in both partonic and hadronic phases as implemented in EPOS-3 %will be investigated in terms of jet shapes 
and will be compared with the JEWEL (recoil OFF) results.  

JEWEL is a Monte Carlo simulation program designed for the study of jet quenching in heavy ion collisions. It interfaces a perturbative final state parton shower with the medium and accounts dynamically for the interaction between jet and medium. Aforementioned, the background medium consists of an ensemble of partons whose phase space distribution and flavor composition are determined by an external medium model \cite{JEWEL_main}. The current version of JEWEL uses a variant of the Bjorken model \cite{JEWEL_main} which describes the boost-invariant longitudinal expansion of an ideal QGP. The shower initiated by the hard scattered parton interacts with the background partons and loses energy through elastic and radiative processes. 

JEWEL can be used in two operational modes: recoil ON and OFF \cite{JEWEL_main}. In the recoil OFF mode, the energy and momentum transferred to the background partons are not taken into account in the final state fragmentation. Whereas, in the recoil ON case \cite{JEWEL_main, KunnawalkamElayavalli:2016lzw}, the recoiling partons are inserted into the strings connecting the parton shower and this significantly improves the description of the jet shape observables in JEWEL after proper background subtraction \cite{JEWEL_main}.

\section{Observables and analysis method}

Even though a large number of jet-shape observables are being used to study the jet-medium interactions, in this work, we study two particular observables called differential jet-shape ($\rho$(r)) \cite{cms1} and the angularity or girth {\it g} \cite{Cunqueiro:2015dmx} .
The differential jet shape ($\rho(r)$) \cite{cms1} describes the radial distribution of the jet transverse momentum density inside the jet cone and is defined as: 

$$\rho(r) ~=~ \frac{1}{\delta r} \frac{1}{N_{jet}} \sum_{jets}\frac{\sum_{tracks\in{[r_a,r_b]}}p_{T}^{track}}{p_{T}^{jet}}  $$

Here, the jet cone is divided into several annuli with radial  width of $\delta r$ and each annular ring has an inner radius of $r_a~=~r-\delta r/2$ and outer radius of $r_b~=~r+\delta r/2$. The  $r~=~\sqrt{{(\phi^{track} - \phi^{jet})}^2 + {(\eta^{track} - \eta^{jet})}^2}$ $\leq$ R is the radial distance of the track from the jet axis. The transverse momenta of the tracks and the reconstructed jet are denoted as $p_{T}^{track}$ and $p_{T}^{jet}$ respectively. In the numerator, the transverse momenta of the charged particles inside one annular ring is summed to estimate the fraction of the reconstructed jet momentum $p_{T}^{jet}$ carried by the particles inside each annulus. The final result is obtained after averaging over the total number jets ($N_{jet}$) under consideration.

The Angularity or girth \cite{Cunqueiro:2015dmx} is defined as:

$$g ~=~ \sum_{i\in jet}\frac{p_{T}^{i}}{p_{T}^{jet}}|\Delta R_{jet}^{i}|$$

where $p_{T}^{i}$ denotes the transverse momentum of i-th constituent of the jet with reconstructed jet momentum $p_{T}^{jet}$ and $\Delta R_{jet}^{i}$ is the distance between i-th constituent and the jet axis in $(\eta,\phi)$ space.

These shape observables provide quantitative description of the radial distribution of the jet energy inside the jet cone  and have been extensively measured by all the experimental collaborations at the LHC \cite{cms1, alice, Cunqueiro:2015dmx, Aad:2011kq, skp1}. In heavy ion collisions, based on these observables, the jet core has been found to be more collimated and harder compared to the pp collisions at the same reconstructed jet energy, accompanied by a broadening of the jet at it's periphery \cite{cms1, cms2}. 

The analysis is performed on the charged jets reconstructed with the sequential $anti-k_T$ jet finding algorithm using Fastjet package \cite{Cacciari:2011ma}. Jets are reconstructed with two different values of resolution parameters R~~=~~0.2  and R~~=~~0.3 for $20 <p_{T,chjet} < 40$ GeV/c. The minimum transverse momentum of the tracks allowed for jet reconstruction is set to 0.15 GeV/c. Tracks are selected within $|\eta|<0.9$ and jets are selected with $|\eta_{jet}|<0.7$ and $|\eta_{jet}|<0.6$ for R~~=~~0.2 and R~~=~~0.3 respectively. The jets having at least one particle with transverse momentum above $p_T>5$ GeV/c are considered to reduce the contribution of the combinatorial jets in the selected jet sample \cite{Krizek:2015afa}\cite{Nattrass:2016vac}.

In this work, we have used pp collisions in JEWEL to represent no medium effect and have been compared with Pb-Pb results from both the models. It should be noted that even though we have compared reconstructed jets in pp and Pb-Pb collisions at the same reconstructed jet p$_T$ range, it is likely that the intial jet partons in Pb-Pb were of higher energy before loosing energy in the medium.

For making the observables on the same footing as of experimental data, we have used JEWEL recoil OFF data without background subtraction and in EPOS, only corona (jet) particles are considered to construct the jet.

\section{Results and Discussions}

\begin{figure}
\begin{center}
\includegraphics[height=8.4 cm, width=8.7 cm]{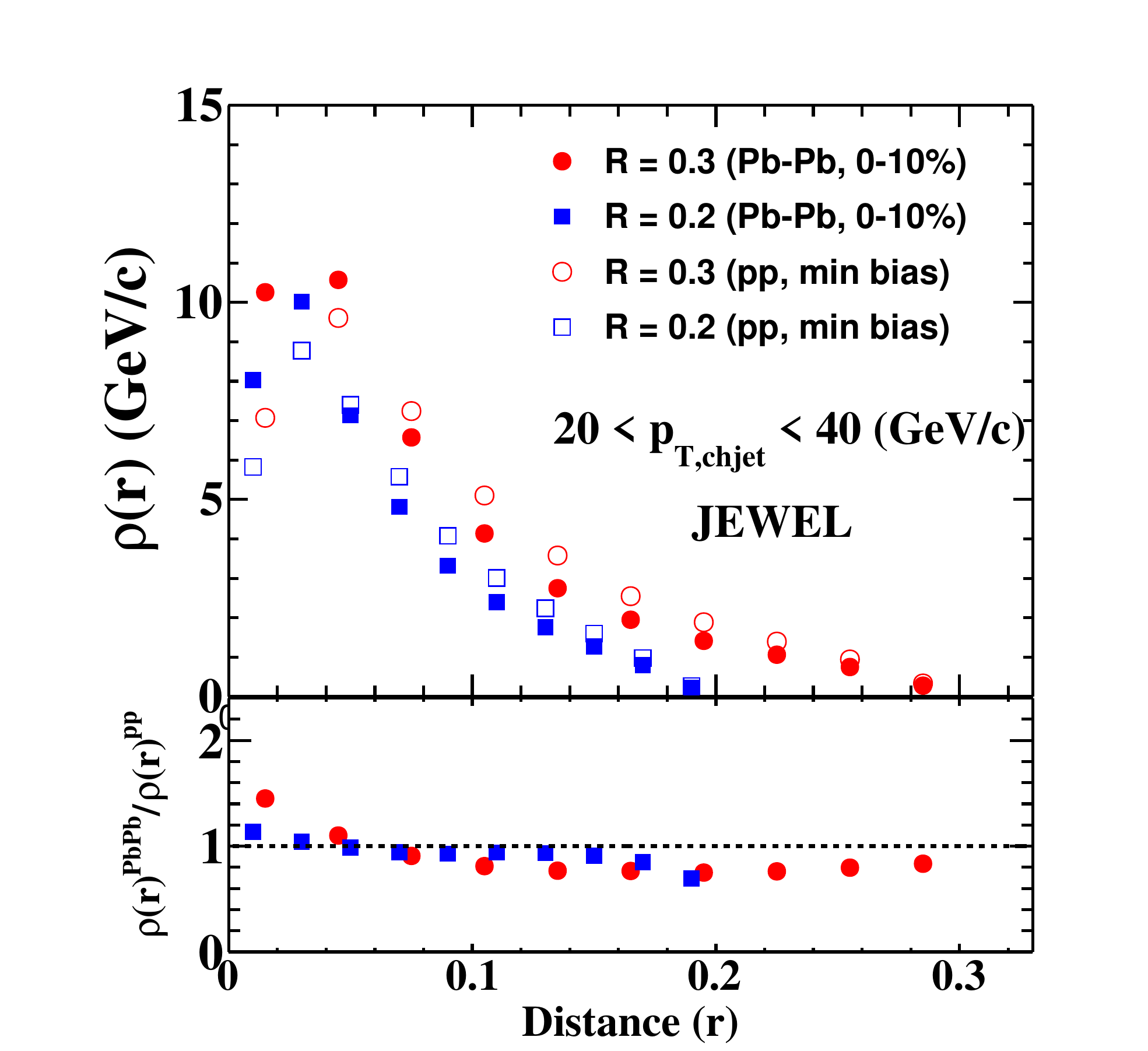}
\caption{[Color online] {\bf Upper panel:} Differential jet shape $\rho(r)$ measured as a function of distance from the jet axis for inclusive charged jets in $20 <p_{T,chjet} < 40$ GeV/c with $R~=~0.2$ and $R~=~0.3$  in 0-10\% central Pb-Pb collisions at $\sqrt{s_{NN}}$ ~=~ 2.76 TeV using the  JEWEL (recoil OFF) event generator and compared with the minimum bias pp results.  {\bf Lower panel:} The jet shape nuclear modification factor, quantified as $\rho(r)^{PbPb}/\rho(r)^{pp}$ }
\label{inclusive-ratioJEWEL}
\end{center}
\end{figure}

\begin{figure}
\begin{center}
\includegraphics[height=8.4 cm, width=8.7 cm]{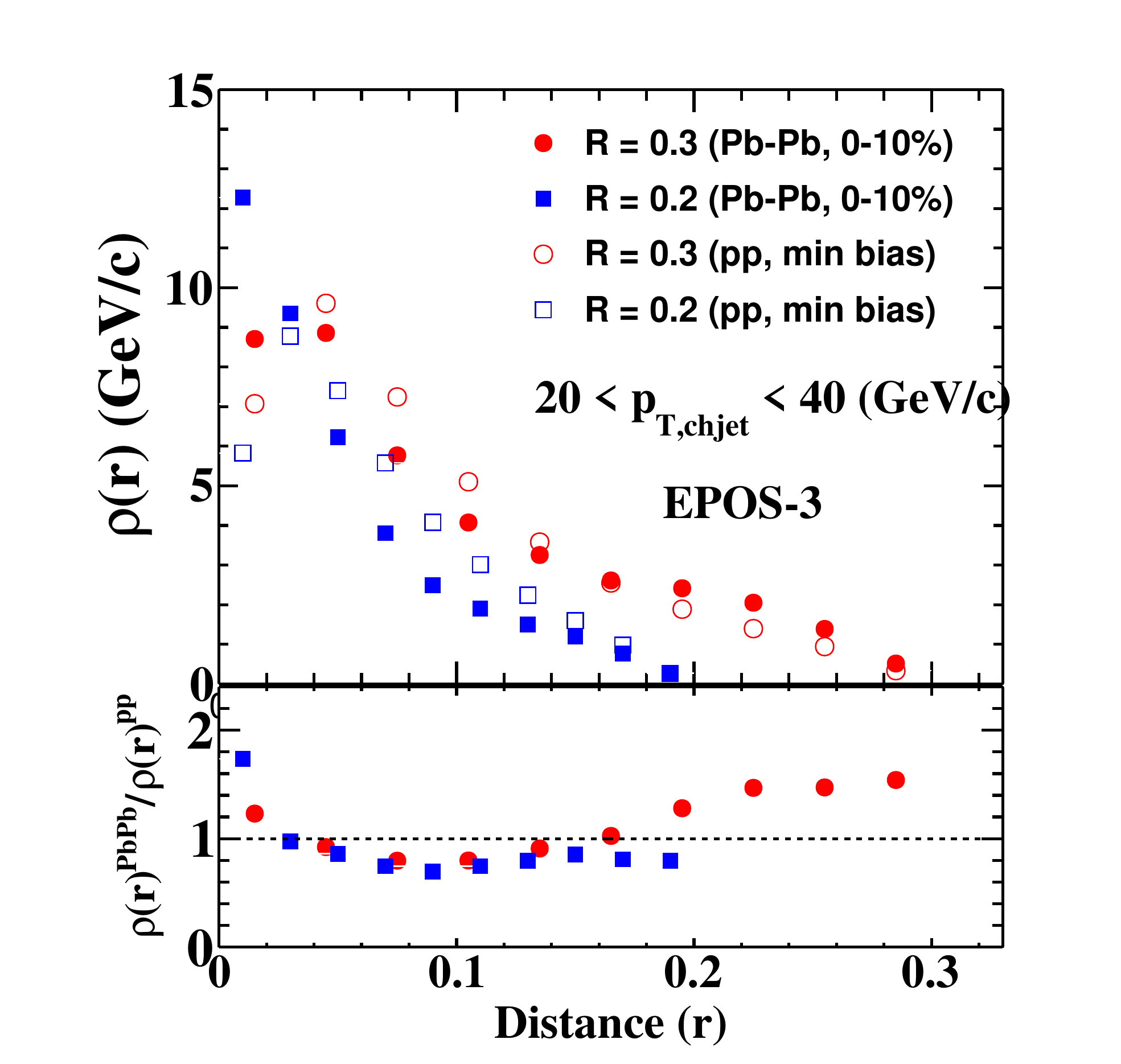}
\caption{[Color online] {\bf Upper panel:} Differential jet shape $\rho(r)$ measured as a function of distance from the jet axis for inclusive charged jets in $20 <p_{T,chjet} < 40$ GeV/c with $R~=~0.2$ and $R~=~0.3$  in 0-10\% central Pb-Pb collisions at $\sqrt{s_{NN}}$ ~=~ 2.76 TeV using the  EPOS-3 event generator and compared with the minimum bias pp results.  {\bf Lower panel:} The jet shape nuclear modification factor, quantified as $\rho(r)^{PbPb}/\rho(r)^{pp}$ }
\label{inclusive-ratioEPOS}
\end{center}
\end{figure}

The top panels of Fig.\ref{inclusive-ratioJEWEL}  and Fig.\ref{inclusive-ratioEPOS} show the comparison of differential jet shape measurements between 0-10\% Pb-Pb collisions and minimum bias pp collisions at 2.76~TeV using the JEWEL (recoil OFF) and EPOS-3 event generators in the jet $p_{T}$ range of 20-40  GeV/c  for R~=~0.2 and R~=~0.3 respectively. 
Qualitatively, we find a set of similarities in results from both models except at high radial distance with resolution parameter R~=~0.3. We can clearly see that for both the models while going radially outward from the jet-axis, the relative difference in $\rho$(r) distribution between 0-10\% Pb-Pb and minimum bias pp collisions changes. This indicates a modification in distribution of energy inside jet cone.

To quantify the medium induced modifications, the jet shape nuclear modification factor ($\rho(r)^{PbPb}/\rho(r)^{pp}$) for both the models are shown at the bottom panels of two figures.
Deviation of the ratio ($\rho(r)^{PbPb}/\rho(r)^{pp}$) from unity indicates a modification to the jet structure in presence of the medium. In comparison to the jet-shape in pp, a narrowing of the jet core (at r $<$  0.02) with higher momentum density has been observed in central Pb-Pb collisions in cases of both EPOS and JEWEL and this is qualitatively similar to the pattern observed in experimental measurements by different collaborations at the LHC \cite{Cunqueiro:2015dmx,cms1,Khachatryan:2016tfj}. 

As discussed earlier, due to in-medium energy loss, jets reconstructed with a fixed R in central Pb-Pb collisions may originate from higher energy initial parton compared to that in pp\cite{Cunqueiro:2015dmx}. 
As the jet core gets harder with increase in jet energy/momentum and is less affected by the presence of a medium, the jet core in central heavy ion collisions can be harder compared to the peripheral and minimum bias pp collisions as shown in Fig.\ref{inclusive-ratioJEWEL} and Fig.\ref{inclusive-ratioEPOS}. Interestingly, the ratio ($\rho(r)^{PbPb}/\rho(r)^{pp}$) becomes less than unity at intermediate radii indicating the in-medium energy loss  in central PbPb collisions.
%but no broadening has been observed at the periphery of jet in case of R~~=~~0.2. 

To understand the redistribution of the lost energy in the medium, we compare the results for two values of the resolution parameter i.e. R~=~0.2 and R~=~0.3 respectively.  Increasing the resolution parameter from R~=~0.2 to R~=~0.3 opens up the possibility of including the energy carried away by softer particles at larger angles from the jet axis. 
%and the measured $\rho(r)$ is shown in Fig.\ref{inclusive-ratioR03}. 
For both the models and the resolution parameters, the ratios ($\rho(r)^{PbPb}/\rho(r)^{pp}$) remain below unity upto a radial distance of 0.2. However at higher radial distances, while it remains below unity for JEWEL (recoil OFF) but goes above unity for EPOS-3 indicating a moderate broadening of jets at the periphery in EPOS-3. 
%Whereas, the JEWEL (recoil OFF) doesn't exhibit any broadening at larger R even in case of R~~=~~0.3. 
The broadening of jets at the periphery in EPOS-3 is qualitatively consistent with the experimental observations which indicate that the energy lost due to jet-medium interaction is distributed at larger distances from the jet axis and represents a clear signature of medium induced modification to the internal jet structure \cite{cms1}. Aforementioned, in EPOS-3, the simple partonic energy  loss mechanism doesn't involve the interaction of jets with hydrodynamically evolving medium. Rather, it depends on the initial geometry of the fireball and local string density along with modified hadronization where the intermediate p$_{T}$ jet string segments pick up hydrodynamically flowing string segments and further interact with the bulk hadrons from the freeze-out. These hard-soft interactions in EPOS-3 contribute to the redistribution of the jet energy inside the jet cone and qualitatively describe the 
jet-shape broadening in heavy ion collisions.

It should however be noted that in case of JEWEL as well, there is a slight trend of increase in the ratio ($\rho(r)^{PbPb}/\rho(r)^{pp}$) towards unity at larger radii. Also, it has been observed \cite{skp2} that at higher jet-p$_T$, the ratio goes above unity at higher radii for JEWEL with recoil ON suggesting that the recoiled medium partons carry and redistribute the energy lost by jets to larger radii.

\begin{figure}
\begin{center}
\includegraphics[height=5.7 cm, width=8.4 cm]{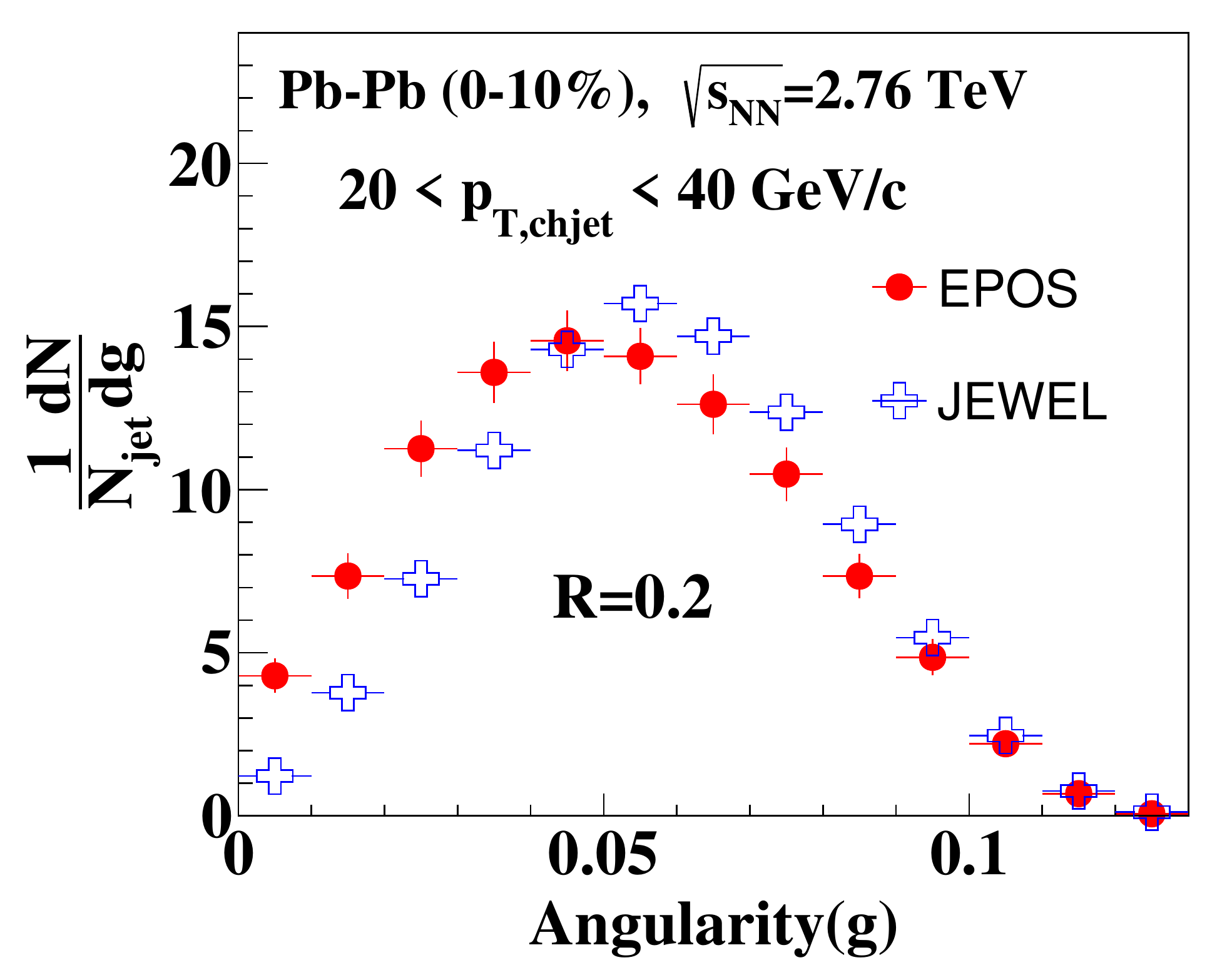}
\caption{[Color online] Angularity (g) measured in 0-10\% central Pb-Pb collisions for inclusive charged jets in $20 <p_{T,chjet} < 40$ GeV/c with $R~=~0.2$ using the EPOS-3 and JEWEL (recoil OFF) event generators.}
\label{inclusive-ratioAng02}
\end{center}
\end{figure}

\begin{figure}
\begin{center}
\includegraphics[height=5.7 cm, width=8.2 cm]{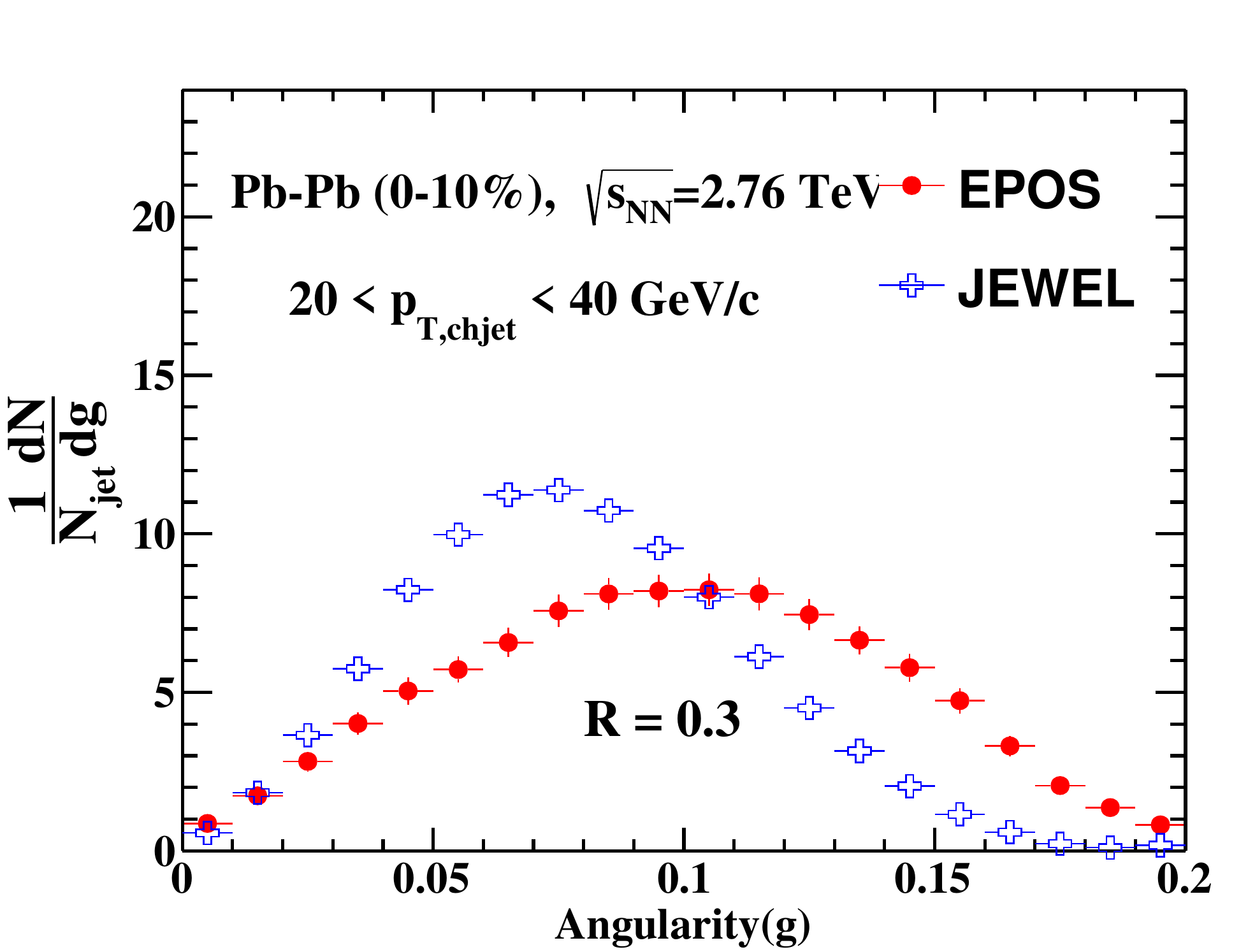}
\caption{[Color online] Same as Fig.\ref{inclusive-ratioAng02} but for jets with $R~=~0.3$ and in the transverse momentum range $20 <p_{T,chjet} < 40$ GeV/c.}
\label{inclusive-ratioAng03}
\end{center}
\end{figure}

To further investigate the radial energy profile of the jets, we also measure the angularity for the two event generators and compared them in Fig.\ref{inclusive-ratioAng02} and Fig.\ref{inclusive-ratioAng03} for R~=~0.2 and R~=~0.3 respectively. For smaller resolution parameter (R~=~0.2), the jet core plays an important role in the measurement of jet shapes and Fig.\ref{inclusive-ratioAng02} indicates that the jet core in EPOS-3 is more collimated than JEWEL. This is qualitatively similar to what we observe in the differential jet shape ($\rho(r)$) measurements as shown in Fig.\ref{inclusive-ratioJEWEL} and Fig.\ref{inclusive-ratioEPOS}. For large radius (R~=~0.3), the medium induced modifications lead to the broadening of the jets at the periphery for both models as shown in Fig.\ref{inclusive-ratioAng03}. The jet in EPOS-3 is harder at  core and broader at  periphery compared to JEWEL and is consistent with the differential jet shape measurements shown earlier. 

%This measurement further indicates that the energy lost by the jets in EPOS-3 appears at larger angles from the jet axis and is in qualitative agreement with the experimentally observed pattern \cite{cms1}.

Our work shows that EPOS-3 which takes into account a simplistic partonic energy loss mechanism and recombination of core and corona string segments at intermediate $p_{T}$ followed by hard-soft re-scatterings in the hadronic phase can qualitatively explain  the jet shape broadening in heavy ion collisions. As discussed earlier, the JEWEL with ``recoil ON" can reasonably explain the jet shapes in heavy ion collisions at higher p$_T$ (where it is easier to make assumptions for background subtractions) thereby emphasizing the role of recoiled medium partons \cite{JEWEL_main}. One of the main differences between the two event generators (EPOS-3 and JEWEL) lies in the way the event is simulated. JEWEL  generates jets and models a medium around it as a collection of scattering centers whose cross sections and distributions in phase space can be chosen from an external medium model. The simplified modeling of the medium in JEWEL (recoil ON and OFF) can't explain the experimentally observed collective behaviors at low $p_{T}$ as well as the nuclear modification factor ($R_{AA}$) for hadrons upto $p_{T}$~=~20 GeV/c in heavy ion collisions \cite{jewel}. In contrast, simulations based on EPOS 3 with a simple partonic energy loss mechanism and secondary hard-soft interactions can describe  the nuclear modification factor \cite{EPOS_main} and in-medium modification to the jet substructure in heavy ion collisions in a consistent way. This observation indicates that the unconventional hard-soft interactions as implemented in EPOS-3  can be instrumental in the realistic modeling of jet-medium interactions.  Further data-model comparisons over a broad range of jet $p_{T}$ with identified constituents will be essential for a better understanding of the medium and the underlying dynamics of the jet-shape broadening in heavy ion collisions.

%due to recombination of jet string segments with hydrodinamically flowing bulk ones along with hard-soft re-scatterings in the hadronic phase
\section{Acknowledgements} 

This material is based upon work supported by the U.S. Department of
Energy Office of Science, Office of Nuclear Physics under Award
Number DE-FG02-92ER-40713. DS would like to thank Federico Ronchetti,  Alessandra Fantoni and Valeria Muccifora for their kind help and support throughout this
work. DS would like to acknowledge the financial support from the CBM-MUCH project grant of BI-IFCC/2016/1082(A). Thanks to Klaus Werner for allowing us to use EPOS-3 for this work. Thanks to Dr. Subikash Chaudhury for useful discussions. Thanks to VECC grid computing team for their 
constant effort to keep the facility running and helping in EPOS and JEWEL  data generation.

\end{document}